\documentclass[submission,copyright,creativecommons]{eptcs}
\providecommand{\event}{SR 2014}

\usepackage{amsmath}
\usepackage{amssymb}
\usepackage{url}
\usepackage{amsthm}
\usepackage{graphicx}
\usepackage[usenames,dvipsnames,svgnames,table]{xcolor}

\newcommand{\ppad}{\textrm{PPAD}}
\newcommand{\fnp}{\textrm{FNP}}
\newcommand{\fp}{\textrm{FP}}

\begin{document}

\theoremstyle{definition}
\newtheorem{theorem}{Theorem}
\newtheorem{definition}[theorem]{Definition}
\newtheorem{problem}[theorem]{Problem}
\newtheorem{assumption}[theorem]{Assumption}
\newtheorem{corollary}[theorem]{Corollary}
\newtheorem{proposition}[theorem]{Proposition}
\newtheorem{lemma}[theorem]{Lemma}
\newtheorem{observation}[theorem]{Observation}
\newtheorem{fact}[theorem]{Fact}
\newtheorem{question}[theorem]{Open Question}
\newtheorem{conjecture}[theorem]{Conjecture}

\newcommand{\red}[1]{{\color{red} #1}}
\newcommand{\blue}[1]{{\color{blue} #1}}
\newcommand{\purple}[1]{{\color{purple} #1}}

\title{Efficient Decomposition of Bimatrix Games (Extended Abstract)\thanks{A full version including proofs omitted here is available as \cite{pauly-xiang-arxiv}.}}


\author{
Xiang Jiang \& Arno Pauly
\institute{Computer Laboratory\\ University of Cambridge, United Kingdom}
\email{Arno.Pauly@cl.cam.ac.uk}
}
\def\titlerunning{Decomposing bimatrix games}
\def\authorrunning{X. Jiang \& A. Pauly}
\maketitle

\begin{abstract}
Exploiting the algebraic structure of the set of bimatrix games, a divide-and-conquer algorithm for finding Nash equilibria is proposed. The algorithm is fixed-parameter tractable with the size of the largest irreducible component of a game as parameter. An implementation of the algorithm is shown to yield a significant performance increase on inputs with small parameters.
\end{abstract}



\section{Introduction}
A bimatrix game is given by two matrices $(A, B)$ of identical dimensions. The first player picks a row $i$, the second player independently picks a column $j$. As a consequence, the first player receives the payoff $A_{ij}$, the second player $B_{ij}$. Both player are allowed to randomize over their choices, and will strive to maximize their expected payoff. A Nash equilibrium is a pair of strategies, such that no player can improve her expected payoff by deviating unilaterally.

If the payoff matrices are given by natural numbers, then there always is a Nash equilibrium using only rational probabilities. The computational task to find a Nash equilibrium of a bimatrix game is complete for the complexity class $\ppad$ \cite{papadimitrioub, daskalakis, denge}. $\ppad$ is contained in $\fnp$, and commonly believed to exceed $\fp$. In particular, it is deemed unlikely that a polynomial-time algorithm for finding Nash equilibria exists.

The next-best algorithmic result to hope for could be a fixed-parameter tractable (fpt) algorithm \cite{downeyfellows}, that is an algorithm running in time $f(k)p(n)$ where $n$ is the size of the game, $p$ a polynomial and $k$ a parameter. For such an algorithm to be useful, the assumption the parameter were usually small needs to be sustainable. The existence of fpt algorithms for finding Nash equilibria with various choices of parameters has been studied in \cite{estivill, hermelin, estivill2}.

In the present paper we demonstrate how \emph{products} and \emph{sums} of games -- and their inverse operations -- can be used to obtain a divide-and-conquer algorithm to find Nash equilibria. This algorithm is fpt, if the size of the largest component not further dividable is chosen as a parameter. \emph{Products} of games were introduced in \cite{paulyincomputabilitynashequilibria} as a means to classify the Weihrauch-degree \cite{brattka3, paulybrattka, paulybrattka3cie, paulykojiro} of finding Nash equilibria for real-valued payoff matrices. \emph{Sums} appear originally in the PhD thesis \cite{paulyphd} of the second author; the algorithm we discuss was implemented in the Bachelor's thesis \cite{jiang} of the first author.

\section{Products and Sums of Games}
Both products and sums admit an intuitive explanation: The product of two games corresponds to playing both games at the same time, while the sum involves playing \emph{matching pennies} to determine which game to play, with one player being rewarded and the other one punished in the case of a failure to agree.

\subsection{Products}
In our definition of products, we let $[ \ , \ ] : \{1, \ldots, n\} \times \{1, \ldots, m\} \to \{1, \ldots, nm\}$ denote the usual bijection $[i, j] = (i - 1)n + j$. The relevant values of $n, m$ will be clear from the context. We point out that $[ \ , \ ]$ is polynomial-time computable and polynomial-time invertible.

\begin{definition}
Given an $n_1 \times m_1$ bimatrix game $(A^1, B^1)$ and an $n_2 \times m_2$ bimatrix game $(A^2, B^2)$, we define the $(n_1n_2) \times (m_1m_2)$ product game $(A^1, B^1) \times (A^2, B^2)$ as $(A, B)$ with $A_{[i_1, i_2][j_1, j_2]} = A_{i_1j_1}^1 + A_{i_2j_2}^2$ and $B_{[i_1, i_2][j_1, j_2]} = B_{i_1j_1}^1 + B_{i_2j_2}^2 $.
\end{definition}

\begin{theorem}
\label{bimatrix:theo:nashproduct1}
If $(x^k, y^k)$ is a Nash equilibrium of $(A^k, B^k)$ for both $k \in \{1, 2\}$, then $(x, y)$ is a Nash equilibrium of $(A, B)$, where $x_{[i_1i_2]} = x_{i_1}^1x_{i_2}^2$ and $y_{[m_1m_2]} = y_{m_1}^1y_{m_2}^2$.
\end{theorem}

\begin{theorem}
\label{bimatrix:theo:nashproduct2}
If $(x, y)$ is a Nash equilibrium of $(A, B)$, then $(x^1, y^1)$ given by $x_{i}^1 = \sum \limits_{l = 1}^{n_2} x_{[i,l]}$ and $y_{j}^1 = \sum \limits_{l = 1}^{m_2} y_{[j,l]}$ is a Nash equilibrium of $(A^1, B^1)$.
\end{theorem}

\subsection{Sums}
The sum of games involves another parameter besides the two component games, which just is a number exceeding the absolute value of all the payoffs.
\begin{definition}
Given an $n_1 \times m_1$ bimatrix game $(A^1, B^1)$ and an $n_2 \times m_2$ bimatrix game $(A^2, B^2)$, we define the $(n_1 + n_2) \times (m_1 + m_2)$ sum game $(A^1, B^1) + (A^2, B^2)$  via the constant $K > \max_{i,j} \{|A_{i,j}|, B_{i,j}|\}$ as $(A, B)$ with: $$A_{i,j} = \begin{cases} A_{ij}^1 & \textnormal{if } i \leq n_1, j \leq m_1 \\ A_{(i-n_1),(j-m_1)}^2 & \textnormal{if } i > n_1, j > m_1 \\ K & \textnormal{otherwise} \end{cases}$$ $$B_{i,j} = \begin{cases} B_{ij}^1 & \textnormal{if } i \leq n_1, j \leq m_1 \\ B_{(i-n_1),(j-m_1)}^2 & \textnormal{if } i > n_1, j > m_1 \\ - K & \textnormal{otherwise} \end{cases}$$
\end{definition}

\begin{lemma}
\label{bimatrix:lemma:summatchingpennies}
Let $(x, y)$ be a Nash equilibrium of $(A^1, B^1) + (A^2, B^2)$. Then $0 < \left ( \sum_{i = 1}^{n_1} x_i \right ) < 1$ and $0 < \left ( \sum_{j = 1}^{m_1} y_j \right ) < 1$.
\end{lemma}

\begin{theorem}
\label{bimatrix:theo:nashsum2}
If $(x, y)$ is a Nash equilibrium of $(A^1, B^1) + (A^2, B^2)$, then a Nash equilibrium $(x^1, y^1)$ of $(A^1, B^1)$ can be obtained as $x^1_i = \frac{x_i}{\sum_{l = 1}^{n_1} x_l}$ and $y^1_j = \frac{y_i}{\sum_{l = 1}^{m_1} y_l}$.
\end{theorem}

\begin{theorem}
\label{bimatrix:theo:nashsum1}
Let $(x^k, y^k)$ be a Nash equilibrium of $(A^k, B^k)$ resulting in payoffs $(P^k, Q^k)$ for both $k \in \{1, 2\}$. Then $(x, y)$ is a Nash equilibrium of $(A^1, B^1) + (A^2, B^2)$, where $x_i = x^1_i \frac{K - Q^2}{2K - Q^1 - Q^2}$ for $i \leq n_1$, $x_i = x^2_{i - n_1} \frac{K - Q^1}{2K - Q^1 - Q^2}$ for $i > n_1$, $y_j = y^1_j \frac{K - P^2}{2K - P^1 - P^2}$ for $j \leq m_1$, $y_j = y^2_{j - m_1} \frac{K - P^1}{2K - P^1 - P^2}$ for $j > m_1$.
\end{theorem}

If a game is iteratively decomposed into sums, the resulting structure corresponds to a Blackwell game \cite{blackwell,martin2} of finite length. The reasoning underlying the theorems above then provides a means of backwards-induction to show that such games always admit Nash equilibria without a direct appeal to their normal form version. The latter observation is the foundation for \cite[Corollary 8]{paulyleroux2-arxiv}.

\section{Examples}
In order to illuminate both how the operations work, and how the component games can be recovered from the compound game, we shall briefly consider some examples. As all relevant features already appear for zero-sum games, we shall restrict the examples to this case, and suppress explicit reference to the second player's payoffs.
\[A := \left ( \begin{array}{cccc} 1 & 2 & 3 & 4 \\ 0 & 1 & 0 & 1\\ 2 & 2 & 2 & 2 \\ 4 & 1 & 2 & 3\end{array}\right ) \quad B := \left ( \begin{array}{ccc} 0 & 0 & 0 \\ 1 & 0 & 1\\ 1 & 2 & 3 \end{array}\right )\]

\[A \times B = \left ( \begin{array}{cccccccccccc} \red{1} & \red{2} & \red{3} & \red{4} & 1 & 2 & 3 & 4 &  1 & 2 & 3 & 4\\ 2 & 3 & 4 & 5 & 1 & 2 & 3 & 4 & 2 & 3 & 4 & 5\\ 2 & 3 & 4 & 5 & 3 & 4 & 5 & 6 & 4& 5 & 6 & 7\\ \purple{0} & \red{1} & \red{0} & \red{1} & \blue{0} & 1 & 0 & 1 & \blue{0} & 1 & 0 & 1 \\ \blue{1} & 2 & 1 & 2 & \blue{0} & 1 & 0 & 1 & \blue{1} & 2 & 1 & 2\\ \blue{1} & 2 & 1 & 2 & \blue{2} & 3 & 2 & 3 & \blue{3} & 4 & 3 & 4 \\ \red{2} & \red{2} & \red{2} & \red{2} & 2 & 2 & 2 & 2 & 2 & 2 & 2 & 2 \\ 3 & 3 & 3 & 3 & 2 & 2 & 2 & 2 & 3 & 3 & 3 & 3\\ 3 & 3 & 3 & 3 & 4 & 4 & 4 & 4 & 5 & 5 & 5 &5 \\ \red{4} & \red{1} & \red{2} & \red{3} & 4 & 1 & 2 & 3 & 4 & 1 & 2 & 3 \\ 5 & 2 & 3 & 4 & 4 & 1 & 2 & 3 & 5 & 2 & 3 & 4 \\ 5 & 2 & 3 & 4 & 6 & 3 & 4 & 5 & 7 & 4 & 5 & 6\end{array} \right ) \quad \begin{array}{rcl} A_{ij} + B_{11} & = & \red{(A \times B)_{3i - 2,j}}\\ B_{ij} + A_{21} & = & \blue{(A\times B)_{3+i,4j-3}}\end{array}\]

As demonstrated by the colour-markings in the product game, if a game is indeed a product game, then the payoffs of the components can be read off from the payoff matrix of the composed game (plus some constant). Indeed, there are many different positions where the component games are found. Verifying that a game indeed is a product requires to ensure the consistency of these answers.

\[A + B = \left ( \begin{array}{ccccccc} \red{1} & \red{2} & \red{3} & \red{4} & K & K & K \\\red{0} & \red{1} & \red{0} & \red{1} & K & K & K\\ \red{2} & \red{2} & \red{2} & \red{2} & K & K & K \\ \red{4} & \red{1} & \red{2} & \red{3} & K & K & K \\ K & K & K & K & \blue{0} & \blue{0} & \blue{0}\\ K & K & K & K & \blue{1} & \blue{0} & \blue{1} \\ K & K & K & K & \blue{1} & \blue{2} & \blue{3}\end{array} \right ) \]

Recovering the component games from a sum is even simpler: The payoff matrices are found in the left-upper and right-lower corner, while the remaining two cornes are covered by a suitable constant $K$. The latter allows us to determine the precise size of the corners.
\section{The algorithm}
\label{sec:algo}
Our basic algorithm proceeds as follows: To solve a game $(A, B)$
\begin{enumerate}
\item test whether $(A, B)$ is the sum of $(A^1, B^1)$ and $(A^2, B^2)$ via some constant $K$. If yes, solve $(A^1, B^1)$ and $(A^2, B^2)$ and combine their Nash equilibria to an equilibrium of $(A, B)$ via Theorem \ref{bimatrix:theo:nashsum1}. If no,
\item test whether $(A, B)$ is the product of $(A^1, B^1)$ and $(A^2, B^2)$. If yes, solve $(A^1, B^1)$ and $(A^2, B^2)$ and combine their Nash equilibria to an equilibrium of $(A, B)$ via Theorem \ref{bimatrix:theo:nashproduct1}. If no,
\item find a Nash equilibrium of $(A, B)$ by some other means.
\end{enumerate}

For some $n \times m$ game $(A, B)$ let let $S(A, B)$ denote its size, i.e.~$S(A, B) = nm$, and let $\lambda(A, B)$ be the size of the largest game for which $3.$ in our algorithm is called. Let $f(k)$ be the time needed for the external algorithm called in $3.$ on a game of size $k$. Then the runtime of our algorithm is bounded by $O(S^{2}f(\lambda))$, in particular, it is an $fpt$-algorithm:

Testing whether a game is a sum, and computing the components, if applicable, can be done in linear time: As the value $K$ has to appear in the corner, one look up whether two suitable rectangular regions have payoffs $K$ and $-K$. If this is the case, the remaining two rectangular regions contain the two subgames, provided they do not contain payoffs $p$ with $|p| \geq K$. The sum of the sizes of the components is less than the size of the original game. Finally, combining Nash equilibria can be done in linear time, too. Only this case for yield quadratic runtime.

Whether a game is a product of factors of a fixed size can also be tested in linear time. Essentially, the payoffs of the putative component games are found as differences between corresponding payoffs in the original game. Then the product of the two component games can be computed, and finally compared to the original game to verify the decomposition. Testing the different possible factors adds an additional factor $\sqrt{S}$ for this part. The product of the sizes of the factors is equal to size of the original game. Again, combining the Nash equilibria takes linear time. The underlying recurrence relation on its own would yield a time bound of $O(S^{1.5}\log S)$, hence the quadratic runtime bound from the first relation is dominating.

Note that a game cannot simultaneously be the result of a sum and a product of smaller games. Thus, a full decomposition (and the size of the largest component) of a game is independent of the order in which decomposability is tested.

As a slight modification of our algorithm, one can eliminate (iteratively) strictly dominated strategies at each stage of the algorithm. We recall that a strategy $i$ of some player is called strictly dominated by some other strategy $j$, if against any strategy chosen by the opponent, $i$ provides its player with a strictly better payoff than $j$. A strictly dominated strategy can never be used in a Nash equilibrium. It is easy to verify that a game decomposable as a sum never has any strictly dominated strategies, but may occur as the result of the elimination of such strategies. Hence, including an elimination step for each stage increases the potential for decomposability.

\begin{proposition}
Elimination of strictly dominated strategies commutes with decomposition of products, i.e.~the reduced form of the product is the product of the reduced forms of the factors.
\end{proposition}
The algorithm remains $fpt$ if such a step is included, using e.g.~the algorithm presented in \cite{knuth}. The exponent would presumably increase to $O(S^4f(\lambda))$ though. Discussion of complexity issues regarding removal of dominated strategies can be found in e.g.~\cite{harrenstein,paulycomplexityofise}.

\section{Empirical evaluation}
Only a small fraction of the bimatrix games of a given size and bounded integer payoffs will be decomposable by our techniques, this limiting the applicability of the algorithm in Section \ref{sec:algo}. In particular, if sample games were drawn from a uniform distribution, one should not expect any speedup using decomposability-tests. However, to some extent we can expect patterns in the definitions of real-world game situation to increase the decomposability of the derived bimatrix games. For example, the structure of Poker-style games implies decomposability, as can be concluded from the considerations\footnote{Making decisions on whether to \emph{fold}, \emph{call} or \emph{raise} corresponds to choosing the type of the remaining game, i.e.~a sum decomposition. Similarly, the cards chosen by chance induce a product decomposition of the expected values of the game.} in \cite{gilpin}. Note that an explicit understanding of such patterns is not required to benefit from our algorithm -- a reasonable expectation that suitable patterns could occur is enough to justify the use of our algorithm, which then identifies the actual patterns.

To obtain a first impression whether using the decomposition algorithm is indeed beneficial for computing Nash equilibria, a collection of 100 random decomposable games was created. Each game has 95-105 strategies per player, and payoff values range from 0 to 50. The decomposability was ensure by creating a random tree representing the relevant decomposition structure first, using probabilities of 0.4 each for sum and product decomposition, and of 0.2 for an elimination of strictly dominated strategies step. The height of the trees was limited to 80, additionally vertices corresponding to games of size up to 6 were turned into leaves. At the leaves, the payoffs were chosen uniformly subject to the constraints derived from the structure and the overall constraint of payoff values being between 0 and 50. Finally, the corresponding bimatrix games were computed. As the payoffs for both players were chosen independently, the expected fraction of zero-sum games in the sample set is negligible.

Both as a benchmark, and in order to compute Nash equilibria of the irreducible component games, the tool \textsc{Gambit} \cite{gambit} was used. \textsc{Gambit} offers a variety of algorithm for computing Nash equilibria of bimatrix games, we used:
\begin{enumerate}
\item gambit-enummixed: using extreme point enumeration
\item gambit-gnm: using a global Newton method approach
\item gambit-lcp: using linear complementarity
\item gambit-simpdiv: using simplicial subdivision
\end{enumerate}

\begin{figure}[htbp]
\centering
\includegraphics[width=0.8\textwidth]{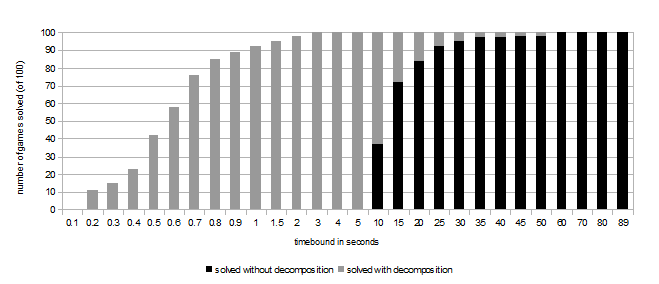}
\caption{gambit-gnm}
\label{gnm}
\end{figure}

\begin{figure}[htbp]
\centering
\includegraphics[width=0.8\textwidth]{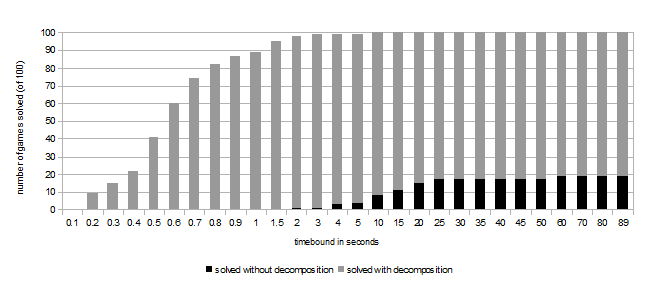}
\caption{gambit-lcp}
\label{lcp}
\end{figure}

Figures 1. \& 2.~show for gambit-gmn and gambit-lcp how many of our decomposable example games could be solved in some given time bound (per game, not total) using only the \textsc{Gambit} algorithm directly, or exploiting decomposition implemented in C++ first. Despite the fact that our decomposition algorithm was not optimized, it turned out that using decomposition almost all games could be solved in under 3 seconds, whereas even gambit-gnm\footnote{Which is based on \cite{govindan}.} as the fastest \textsc{Gambit} algorithm on the sample took 30 seconds for a similar feat. Neither gambit-enummixed nor gambit-subdiv were able to solve any of the example games in less than 90 seconds without decomposition. Note that the \textsc{Gambit} algorithms generally include elimination of strictly dominated strategies as well, so this alone cannot account for the differences. Thus, there is clear indication that on suitable data, exploiting the algebraic structure underlying the decomposition algorithm yields a significant increase in performance.
\bibliographystyle{eptcs}
\bibliography{../spieltheorie}

\begin{thebibliography}{10}
\providecommand{\bibitemdeclare}[2]{}
\providecommand{\surnamestart}{}
\providecommand{\surnameend}{}
\providecommand{\urlprefix}{Available at }
\providecommand{\url}[1]{\texttt{#1}}
\providecommand{\href}[2]{\texttt{#2}}
\providecommand{\urlalt}[2]{\href{#1}{#2}}
\providecommand{\doi}[1]{doi:\urlalt{http://dx.doi.org/#1}{#1}}
\providecommand{\bibinfo}[2]{#2}

\bibitemdeclare{article}{blackwell}
\bibitem{blackwell}
\bibinfo{author}{D.~\surnamestart Blackwell\surnameend} (\bibinfo{year}{1969}):
  \emph{\bibinfo{title}{Infinite {$G_\delta$} games with imperfect
  information}}.
\newblock {\sl \bibinfo{journal}{Zastowania Matematyki Applicationes
  Mathematicae}}.

\bibitemdeclare{article}{harrenstein}
\bibitem{harrenstein}
\bibinfo{author}{Felix \surnamestart Brandt\surnameend},
  \bibinfo{author}{Markus \surnamestart Brill\surnameend},
  \bibinfo{author}{Felix \surnamestart Fischer\surnameend} \&
  \bibinfo{author}{Paul \surnamestart Harrenstein\surnameend}
  (\bibinfo{year}{2011}): \emph{\bibinfo{title}{On the Complexity of Iterated
  Weak Dominance in Constant-sum Games}}.
\newblock {\sl \bibinfo{journal}{Theory of Computing Systems}}
  \bibinfo{volume}{49}(\bibinfo{number}{1}), pp. \bibinfo{pages}{162--181}.
\newblock \urlprefix\url{http://dx.doi.org/10.1007/s00224-010-9282-7}.

\bibitemdeclare{article}{paulybrattka}
\bibitem{paulybrattka}
\bibinfo{author}{Vasco \surnamestart Brattka\surnameend},
  \bibinfo{author}{Matthew \surnamestart de~Brecht\surnameend} \&
  \bibinfo{author}{Arno \surnamestart Pauly\surnameend} (\bibinfo{year}{2012}):
  \emph{\bibinfo{title}{Closed Choice and a Uniform Low Basis Theorem}}.
\newblock {\sl \bibinfo{journal}{Annals of Pure and Applied Logic}}
  \bibinfo{volume}{163}(\bibinfo{number}{8}), pp. \bibinfo{pages}{968--1008},
  \doi{10.1016/j.apal.2011.12.020}.

\bibitemdeclare{article}{brattka3}
\bibitem{brattka3}
\bibinfo{author}{Vasco \surnamestart Brattka\surnameend} \&
  \bibinfo{author}{Guido \surnamestart Gherardi\surnameend}
  (\bibinfo{year}{2011}): \emph{\bibinfo{title}{Effective Choice and
  Boundedness Principles in Computable Analysis}}.
\newblock {\sl \bibinfo{journal}{Bulletin of Symbolic Logic}}
  \bibinfo{volume}{1}, pp. \bibinfo{pages}{73 -- 117},
  \doi{10.2178/bsl/1294186663}.
\newblock \bibinfo{note}{ArXiv:0905.4685}.

\bibitemdeclare{incollection}{paulybrattka3cie}
\bibitem{paulybrattka3cie}
\bibinfo{author}{Vasco \surnamestart Brattka\surnameend},
  \bibinfo{author}{St\'ephane \surnamestart Le~Roux\surnameend} \&
  \bibinfo{author}{Arno \surnamestart Pauly\surnameend} (\bibinfo{year}{2012}):
  \emph{\bibinfo{title}{On the Computational Content of the {B}rouwer Fixed
  Point Theorem}}.
\newblock In \bibinfo{editor}{S.Barry \surnamestart Cooper\surnameend},
  \bibinfo{editor}{Anuj \surnamestart Dawar\surnameend} \&
  \bibinfo{editor}{Benedikt \surnamestart L\"owe\surnameend}, editors: {\sl
  \bibinfo{booktitle}{How the World Computes}}, {\sl \bibinfo{series}{Lecture
  Notes in Computer Science}} \bibinfo{volume}{7318},
  \bibinfo{publisher}{Springer Berlin Heidelberg}, pp. \bibinfo{pages}{56--67},
  \doi{10.1007/978-3-642-30870-3\_7}.

\bibitemdeclare{article}{denge}
\bibitem{denge}
\bibinfo{author}{Xi~\surnamestart Chen\surnameend}, \bibinfo{author}{Xiaotie
  \surnamestart Deng\surnameend} \& \bibinfo{author}{Shang-Hua \surnamestart
  Teng\surnameend} (\bibinfo{year}{2009}): \emph{\bibinfo{title}{Settling the
  Complexity of Computing Two-player Nash Equilibria}}.
\newblock {\sl \bibinfo{journal}{J. ACM}}
  \bibinfo{volume}{56}(\bibinfo{number}{3}), pp. \bibinfo{pages}{14:1--14:57},
  \doi{10.1145/1516512.1516516}.

\bibitemdeclare{incollection}{daskalakis}
\bibitem{daskalakis}
\bibinfo{author}{Constantinos \surnamestart Daskalakis\surnameend},
  \bibinfo{author}{Paul \surnamestart Goldberg\surnameend} \&
  \bibinfo{author}{Christos \surnamestart Papadimitriou\surnameend}
  (\bibinfo{year}{2006}): \emph{\bibinfo{title}{The Complexity of Computing a
  {N}ash Equilibrium}}.
\newblock In: {\sl \bibinfo{booktitle}{38th ACM Symposium on Theory of
  Computing}}, pp. \bibinfo{pages}{71--78}, \doi{10.1145/1132516.1132527}.

\bibitemdeclare{book}{downeyfellows}
\bibitem{downeyfellows}
\bibinfo{author}{Rod \surnamestart Downey\surnameend} \&
  \bibinfo{author}{Michael \surnamestart Fellows\surnameend}
  (\bibinfo{year}{1999}): \emph{\bibinfo{title}{Parameterized Complexity}}.
\newblock \bibinfo{publisher}{Springer}, \doi{10.1007/978-1-4612-0515-9}.

\bibitemdeclare{inproceedings}{estivill}
\bibitem{estivill}
\bibinfo{author}{Vladimir \surnamestart Estivill-Castro\surnameend} \&
  \bibinfo{author}{Mahdi \surnamestart Parsa\surnameend}
  (\bibinfo{year}{2009}): \emph{\bibinfo{title}{Computing {N}ash equilibria
  Gets Harder {--} New Results Show Hardness Even for Parameterized
  Complexity}}.
\newblock In \bibinfo{editor}{Rod \surnamestart Downey\surnameend} \&
  \bibinfo{editor}{Prabhu \surnamestart Manyem\surnameend}, editors: {\sl
  \bibinfo{booktitle}{CATS 2009}}, {\sl
  \bibinfo{series}{CRPIT}}~\bibinfo{volume}{94}.

\bibitemdeclare{incollection}{estivill2}
\bibitem{estivill2}
\bibinfo{author}{Vladimir \surnamestart Estivill-Castro\surnameend} \&
  \bibinfo{author}{Mahdi \surnamestart Parsa\surnameend}
  (\bibinfo{year}{2011}): \emph{\bibinfo{title}{Single Parameter FPT-Algorithms
  for Non-trivial Games}}.
\newblock In \bibinfo{editor}{Costas \surnamestart Iliopoulos\surnameend} \&
  \bibinfo{editor}{William \surnamestart Smyth\surnameend}, editors: {\sl
  \bibinfo{booktitle}{Combinatorial Algorithms}}, {\sl \bibinfo{series}{Lecture
  Notes in Computer Science}} \bibinfo{volume}{6460},
  \bibinfo{publisher}{Springer}, pp. \bibinfo{pages}{121--124}.
\newblock \urlprefix\url{http://dx.doi.org/10.1007/978-3-642-19222-7_13}.

\bibitemdeclare{inproceedings}{gilpin}
\bibitem{gilpin}
\bibinfo{author}{Andrew \surnamestart Gilpin\surnameend},
  \bibinfo{author}{Javier \surnamestart Pena\surnameend},
  \bibinfo{author}{Samid \surnamestart Hoda\surnameend} \&
  \bibinfo{author}{Tuomas \surnamestart Sandholm\surnameend}
  (\bibinfo{year}{2007}): \emph{\bibinfo{title}{Gradient-based algorithms for
  finding {N}ash equilibria in extensive form games}}.
\newblock In: {\sl \bibinfo{booktitle}{Proceedings of the 18th Int Conf on Game
  Theory}}, \doi{10.1007/978-3-540-77105-0\_9}.

\bibitemdeclare{article}{govindan}
\bibitem{govindan}
\bibinfo{author}{Srihari \surnamestart Govindan\surnameend} \&
  \bibinfo{author}{Robert \surnamestart Wilson\surnameend}
  (\bibinfo{year}{2003}): \emph{\bibinfo{title}{A Global Newton Method to
  Compute Nash Equilibria}}.
\newblock {\sl \bibinfo{journal}{Journal of Economic Theory}}
  \bibinfo{volume}{110}(\bibinfo{number}{1}), pp. \bibinfo{pages}{65--86},
  \doi{10.1016/S0022-0531(03)00005-X}.

\bibitemdeclare{article}{hermelin}
\bibitem{hermelin}
\bibinfo{author}{Danny \surnamestart Hermelin\surnameend},
  \bibinfo{author}{Chien-Chung \surnamestart Huang\surnameend},
  \bibinfo{author}{Stefan \surnamestart Kratsch\surnameend} \&
  \bibinfo{author}{Magnus \surnamestart Wahlstr{\"o}m\surnameend}
  (\bibinfo{year}{2010}): \emph{\bibinfo{title}{Parameterized Two-Player Nash
  Equilibrium}}.
\newblock {\sl \bibinfo{journal}{CoRR}} \bibinfo{volume}{abs/1006.2063}.
\newblock \urlprefix\url{http://arxiv.org/abs/1006.2063}.

\bibitemdeclare{article}{paulykojiro}
\bibitem{paulykojiro}
\bibinfo{author}{Kojiro \surnamestart Higuchi\surnameend} \&
  \bibinfo{author}{Arno \surnamestart Pauly\surnameend} (\bibinfo{year}{2013}):
  \emph{\bibinfo{title}{The degree-structure of {W}eihrauch-reducibility}}.
\newblock {\sl \bibinfo{journal}{Logical Methods in Computer Science}}
  \bibinfo{volume}{9}(\bibinfo{number}{2}), \doi{10.2168/LMCS-9(2:2)2013}.

\bibitemdeclare{phdthesis}{jiang}
\bibitem{jiang}
\bibinfo{author}{Xiang \surnamestart Jiang\surnameend} (\bibinfo{year}{2011}):
  \emph{\bibinfo{title}{Efficient Decomposition of Games}}.
\newblock \bibinfo{type}{Bachelor's thesis}, \bibinfo{school}{University of
  Cambridge}.

\bibitemdeclare{misc}{pauly-xiang-arxiv}
\bibitem{pauly-xiang-arxiv}
\bibinfo{author}{Xiang \surnamestart Jiang\surnameend} \& \bibinfo{author}{Arno
  \surnamestart Pauly\surnameend} (\bibinfo{year}{2012}):
  \emph{\bibinfo{title}{Efficient Decomposition of Bimatrix Games}}.
\newblock \bibinfo{howpublished}{http://arxiv.org/abs/1212.6355}.

\bibitemdeclare{article}{knuth}
\bibitem{knuth}
\bibinfo{author}{Donald \surnamestart Knuth\surnameend},
  \bibinfo{author}{Christos \surnamestart Papadimitriou\surnameend} \&
  \bibinfo{author}{John \surnamestart Tsitsiklis\surnameend}
  (\bibinfo{year}{1988}): \emph{\bibinfo{title}{A note on strategy elimination
  in bimatrix games}}.
\newblock {\sl \bibinfo{journal}{Operations Research Letters}}
  \bibinfo{volume}{7}(\bibinfo{number}{3}), pp. \bibinfo{pages}{103--107},
  \doi{10.1016/0167-6377(88)90075-2}.

\bibitemdeclare{misc}{paulyleroux2-arxiv}
\bibitem{paulyleroux2-arxiv}
\bibinfo{author}{St\'ephane \surnamestart Le~Roux\surnameend} \&
  \bibinfo{author}{Arno \surnamestart Pauly\surnameend} (\bibinfo{year}{2014}):
  \emph{\bibinfo{title}{Infinite sequential games with real-valued payoffs}}.
\newblock \bibinfo{howpublished}{arXiv:1401.3325}.

\bibitemdeclare{article}{martin2}
\bibitem{martin2}
\bibinfo{author}{Donald~A. \surnamestart Martin\surnameend}
  (\bibinfo{year}{1998}): \emph{\bibinfo{title}{The Determinacy of Blackwell
  Games}}.
\newblock {\sl \bibinfo{journal}{Journal of Symbolic Logic}}
  \bibinfo{volume}{63}(\bibinfo{number}{4}), pp. \bibinfo{pages}{1565--1581},
  \doi{10.2307/2586667}.

\bibitemdeclare{misc}{gambit}
\bibitem{gambit}
\bibinfo{author}{Richard \surnamestart McKelvey\surnameend},
  \bibinfo{author}{Andrew \surnamestart McLennan\surnameend} \&
  \bibinfo{author}{Theodore \surnamestart Turocy\surnameend}
  (\bibinfo{year}{2010}): \emph{\bibinfo{title}{Gambit: Software Tools for Game
  Theory}}.
\newblock \bibinfo{howpublished}{http://www.gambit-project.org}.
\newblock \bibinfo{note}{Version 0.2010.09.01}.

\bibitemdeclare{article}{papadimitrioub}
\bibitem{papadimitrioub}
\bibinfo{author}{Christos~H. \surnamestart Papadimitriou\surnameend}
  (\bibinfo{year}{1994}): \emph{\bibinfo{title}{On the complexity of the parity
  argument and other inefficient proofs of existence}}.
\newblock {\sl \bibinfo{journal}{Journal of Computer and Systems Science}}
  \bibinfo{volume}{48}(\bibinfo{number}{3}), pp. \bibinfo{pages}{498--532},
  \doi{10.1016/S0022-0000(05)80063-7}.

\bibitemdeclare{misc}{paulycomplexityofise}
\bibitem{paulycomplexityofise}
\bibinfo{author}{Arno \surnamestart Pauly\surnameend} (\bibinfo{year}{2009}):
  \emph{\bibinfo{title}{The Complexity of Iterated Strategy Elimination}}.
\newblock \bibinfo{howpublished}{arXiv:0910.5107}.

\bibitemdeclare{article}{paulyincomputabilitynashequilibria}
\bibitem{paulyincomputabilitynashequilibria}
\bibinfo{author}{Arno \surnamestart Pauly\surnameend} (\bibinfo{year}{2010}):
  \emph{\bibinfo{title}{How Incomputable is Finding {N}ash Equilibria?}}
\newblock {\sl \bibinfo{journal}{Journal of Universal Computer Science}}
  \bibinfo{volume}{16}(\bibinfo{number}{18}), pp. \bibinfo{pages}{2686--2710},
  \doi{10.3217/jucs-016-18-2686}.

\bibitemdeclare{phdthesis}{paulyphd}
\bibitem{paulyphd}
\bibinfo{author}{Arno \surnamestart Pauly\surnameend} (\bibinfo{year}{2012}):
  \emph{\bibinfo{title}{Computable Metamathematics and its Application to Game
  Theory}}.
\newblock Ph.D. thesis, \bibinfo{school}{University of Cambridge}.

\end{thebibliography}

\section*{Acknowledgements}
We are grateful for various helpful comments from referees. One comment in particular was instrumental in improving the runtime analysis of our algorithm to a quadratic exponent.
\end{document}